\documentclass[a4paper]{jpconf}

\usepackage{amsmath,amssymb}
\usepackage{epsfig}
\usepackage[square,sort&compress]{natbib}
\usepackage{graphicx}
\usepackage{float}
\usepackage{array}
\usepackage{rotating}
\usepackage{bm}
\usepackage{url}

\begin{document}
\title{Noise in pulsar timing arrays}

\author{Yan Wang}

\address{School of Physics, Huazhong University of Science and Technology, 1037 Luoyu Road, Wuhan, Hubei Province 430074, China}

\address{Center for Advanced Radio Astronomy, University of Texas at Brownsville, 1 West University Boulevard, Brownsville, Texas 78520, USA}

\ead{ywang12@hust.edu.cn}

\begin{abstract}
To successfully detect gravitational waves with pulsar timing arrays, 
we need to have a comprehensive understanding of the physical origins and statistical characteristics 
of the noise in pulse arrival times and identify mitigation methods 
to reduce the noise. In this paper we will review radiometer noise, phase jitter 
noise and timing noise in the noise budget of pulsar timing and show various efforts 
used to reduce them. We will briefly discuss the results of 
an overall assessment of the components and physical causes of the timing 
residuals for millisecond pulsars in the North American Nanohertz Observatory 
for Gravitational Waves (NANOGrav). 
\end{abstract}

\section{Introduction}

Pulsar timing arrays (PTAs) are striving to detect very low frequency 
($10^{-9}-10^{-6}$ Hz) gravitational waves (GWs) by observing a set of 
extremely stable millisecond pulsars (MSPs). Detection can be achieved by 
observing 20--40 pulsars over 5--10 years, assuming monthly observation 
cadence and 100 ns for root mean square (RMS) of timing residuals which 
is presumably dictated by white Gaussian 
noise \citep{2005ApJ...625L.123J}. However, detection can be delayed by 
up to about 10 years depending on the level of red noise and the possible 
errors (e.g., ephemeris, polarization calibration, time transfer) in the 
highest timing precision \citep{2013CQGra..30v4015S}. 
So far, upper limits have been reported for the amplitude of the stochastic 
GW background \citep{2011MNRAS.414.3117V, 2013ApJ...762...94D, 2013Sci...342..334S} 
and for continuous GW sources \citep{2014ApJ...794..141A, 2014MNRAS.444.3709Z} 
by three major PTAs (NANOGrav \citep{2013ApJ...762...94D}, 
PPTA \citep{2013PASA...30...17M} and EPTA \citep{2010CQGra..27h4014F}). 
Results from the International Pulsar Timing Array (IPTA) \citep{2010CQGra..27h4013H, 
2013CQGra..30v4010M, 2014arXiv1409.4579M} 
which combine the data sets from all PTAs will also become 
available in the near future. 

As in any gravitational wave detection experiment (e.g., LIGO \citep{2009RPPh...72g6901A}, 
Virgo \citep{2011CQGra..28k4002A}, eLISA \citep{2013arXiv1305.5720C}), 
noise characterization and mitigation are the central issues that need to be addressed 
in order to confidently detect and characterize the relatively weak GW signals, and 
carry out detailed astrophysical interpretations of them. For PTAs, a comprehensive 
understanding of the noise error budget of pulse arrival time is crucial to guide us 
to improve instruments, design experiments, and develop algorithms.

To achieve this goal, we need to build a complete measurement model that accounts for the end 
to end errors in the highest timing precision. Here, one end is a rotating 
pulsar, which produces regular radio pulses. The spacially coherent radiation 
carrying these pulses will be distorted by turbulent plasma in interstellar medium (ISM), causing 
dispersion, scattering, refraction, diffraction, etc. of the radio waves 
\citep{2004hpa..book.....L}, along 
the line of propagation to the other end, a telescope. The faint EM signal is 
subsequently collected and focused by the large reflector, received by the 
radiometer and recorded by the backend system at the observatory. From emission 
to reception, there are various errors that can be introduced into the 
final estimation of pulse time of arrivals (TOAs). 

Here, we shall not discuss the propagation effects pertaining to the interstellar 
medium \citep{1977ARA&A..15..479R, 1990ARA&A..28..561R}, since it 
is covered in the account by L. Levin in this volume. Instead, we focus on the 
other errors such as radiometer noise, phase jitter noise and timing 
noise, which are currently under intensive scrutiny of the pulsar timing 
community \citep{2010ApJ...725.1607S, 2014ApJ...794...21D, 2014MNRAS.443.1463S}. 
The rest of this paper is organized as follows. In Sec.~2 we provide an overview of 
these noise sources and methods of mitigation, and in Sec.~3 we discuss briefly the 
efforts and results in the NANOGrav collaboration on noise analysis. The paper is concluded in Sec.~4.

\section{Noise}\label{sec:noise}

In timing analysis, it is a common practice to transform TOAs measured in the topocentric 
reference frame centered at a telescope to the inertial reference frame centered at the Solar 
System Barycenter \citep{2004hpa..book.....L, 2006MNRAS.372.1549E}. 
This transformation includes terms representing deterministic effects such as 
clock correction, time delay due to interstellar medium dispersion, geometric 
time delay (R$\ddot{o}$mer delay), relativistic time delay (Shapiro delay, Einstein 
delay), etc. In addition, various error terms should 
be also added into the time transformation. A comprehensive list of timing errors 
can be found in Table 1 of \citep{2010arXiv1010.3785C}.

In general, we can classify the errors into the ones pertinent to the time tagging 
of pulses (i.e.~TOA measurement by template fitting) and the ones pertinent to the physical properties of 
pulsar or ISM. An example of the latter is the timing of a pulse with infinite SNR and known profile which can be measured 
to an arbitrary precision (no time tagging error), but the irregularity of pulsar rotation or the stochastic 
fluctuation of dispersion measure may introduce additional random components. 
The error can be achromatic which means its influence is independent of the observation 
frequency (timing noise), weakly chromatic (radiometer noise, jitter noise), 
or strongly chromatic (effects rooted from interstellar medium). 


The power spectrum of the noise can be white or red. The TOA fluctuations caused by 
the stochastic GW background have red spectra for individual pulsars, which are angularly 
correlated between pairs of pulsars \cite{1983ApJ...265L..39H}. Thus, for detecting 
this background, it is imperative to assess and reduce, if possible, the confusing 
red noise components rooted from other sources, for example, pulsar and ISM.

\subsection{Radiometer noise}

Radiometer noise is instrumental in origin (thermal electron 
fluctuation) combining the contribution from the sky background 
(dominated by the synchrotron-radiating electrons in the plane of the 
Galaxy). Radiometer noise with a Gaussian probability density function
is additive to the pulse profile, of any of the Stokes parameters. Usually, 
the Stokes I (intensity) is used to measure the TOAs. Radiometer noise is weakly 
chromatic if the declination of pulse flux density with increasing of 
observation frequency cancels out much of the variation of the sky 
background temperature.\footnote{It could be strongly chromatic if 
the pulsar spectrum is flat in the observation band.} 
The resulting TOA is estimated from template fitting of 
integrated pulse profile with theoretical or integrated template. 
The minimum RMS error for TOA estimation due to finite SNR and 
sampling rate is \citep{2010arXiv1010.3785C}: 
\begin{align}\label{eq:raderror1}
\sigma_{\text{SNR}} &= 1\,\mu s \left(\frac{W}{1\,\text{ms}}\right) \left(\frac{N}{10^6}\right)^{-1/2}\left(\frac{1}{\text{SNR}_{1}}\right)\left(\frac{\Delta}{W}\right)^{1/2} \,, \\
&= 0.71 \mu s \left(\frac{W}{1\,\text{ms}}\right)^{3/2}\left(\frac{P}{1\,\text{ms}}\right)^{-1}
\left(\frac{f}{1.4\,\text{GHz}}\right)^{-\alpha}\left(\frac{\Delta f}{1 \, \text{GHz}}\right)^{-1/2}
\left(\frac{N}{10^6}\right)^{-1/2}\left(\frac{S_{\text{sys}}}{S_{1400}}\right) \,.
\end{align}
Here $W$ is the effective pulse width which equals to $W_{\text{FWHM}}/\sqrt{2\pi\ln2}$ 
for a Gaussian pulse profile, $W_{\text{FWHM}}$ is the full width at half maximum of the pulse, 
$N$ is the number of pulse averaged synchronously 
to yield an integrated profile, $\text{SNR}_{1}$ is the single pulse SNR, 
$\Delta$ is the sampling interval, and $P$ is the spin period. Radio 
frequency $f$ and bandwidth $\Delta f$ 
are in GHz, $S_{\text{sys}}$ and $S_{1400}$ are the system equivalent flux 
density in Jy and mean flux density of the pulsar at $1.4$ GHz in mJy. 
Note that optimal result can be archived by increasing the bandwidth and 
integration time, and choosing bright pulsars with short duty cycle. 
The second equality is obtained by using the radiometer equation \citep{2009tra..book.....W}: 
\begin{equation}\label{eq:radiometer}
\Delta T_{\text{sys}}=\frac{T_{\text{sys}}}{\sqrt{n_\text{p} t \Delta f}}  \,,
\end{equation}
and assuming that the flux density of pulsar follows a power law with 
spectral index $\alpha$. $T_{\text{sys}}$ and $\Delta T_{\text{sys}}$ is 
the system temperature of the receiver and its RMS fluctuation, $n_\text{p}$ 
is the number of polarization, and $t$ is the integration time.

Eq. \ref{eq:raderror1}-2 are obtained under the assumption that there is no 
variance for the integrated pulse template over the frequency band, i.e. the 
profile has the same shape at high frequency as at low frequency. However, 
the profile evolution can arise from phenomena intrinsic to pulsar emission 
beam or ISM (DM change, scattering and scintillation). Ignoring it will 
degrade the timing precision promised by the modern broad band receivers 
and backends (see Table \ref{tab:telescopes}).  This so called ``the 
large-bandwidth problem" \cite{2013CQGra..30v4001L} is demonstrated in a 
recent 24-hour global observation for the millisecond PSR 1713+0747 
(c.f. Fig. 4 in \cite{2014ApJ...794...21D}). Current treatment of this problem 
is to generate templates for each frequency channel, add additional fitting 
parameters (JUMPs) to handle the possible offsets between them, and allow the 
standard TOA analysis packages such as \texttt{tempo} and \texttt{tempo2} 
to find out the best fit value for them \citep{2013ApJ...762...94D}. 
A more consistent and efficient method may be to create a two-dimensional pulse portrait 
(rather than one-dimensional profiles at different frequencies) which 
takes into account of differential profile evolution and time offset 
as a smooth function of frequency. This strategy has been explored for broad band 
data in \cite{2014ApJ...790...93P, 2014MNRAS.443.3752L}.

\begin{table}[h] 
\caption{\label{blobs} Wide band receivers and backends currently used 
or planned for pulsar observation.}
\begin{center}\label{tab:telescopes}
\begin{tabular}{llllllll}
\br
Telescope & D(m)& Receiver & $f$(GHz) & $T_{\text{sys}}$ & Backend & $\Delta f$(MHz) & Ref. \\
\mr
GBT         & 100 & L-band & 1.15-1.73 & 20 & GUPPI & 800 & \citep{2014GBTguide} \\
Arecibo     & 305 & L-band & 1.15-1.73 & 25 & PUPPI & 800 & \citep{AreciboLw,AreciboPUPPI} \\  
Parkes      & 64 & Multibeam & 1.23-1.53 & 28 & APSR &1000 & \cite{2013PASA...30...17M} \\ 
Effelsberg  & 100 & UBB & 0.6-3 & 24 & ASTERIX & 512 & \citep{EffUBB,Effasterix} \\ 
FAST        & 500 & L-band & 1.15-1.72 & 25 & -- & -- & \cite{2011IJMPD..20..989N} \\ 
\br
\end{tabular}
\end{center}
\end{table}

\subsection{Pulse phase jitter and amplitude modulation noise}

Increasing the SNR of the integrated profile will reduce the radiometer noise, however it is 
not the only justification for using integrated pulse profile in 
TOA measurement. Individual pulse usually jitters in phase at the level 
of single pulse width, and its amplitude can change more than 100\% from 
pulse to pulse (c.f. Fig.~4 of \cite{2002ApJ...564..333M} for 
PSR J1740+1000). Thus measuring individual pulses even with large 
$\text{SNR}_1$ will result in a TOA uncertainty in the order of the pulse width. 
Phase jitter and amplitude modulation are related to the stability of 
the integrated pulse shape which is determined by the shape of 
individual pulse and the probability density function of the phase 
jitter. A stable pulse profile can be obtained by summing over at least 
several hundreds of individual pulses, and the associated TOA uncertainty due 
to jitter roughly scales inversely as square root of the number of 
the individual pulses \citep{1995ApJ...452..814R}. 

Pulse phase jitter and amplitude modulation appear in all well 
studied pulsars. It is weakly chromatic. 
Jitter noise is not additive in nature as the radiometer noise. In fact, it changes the 
integrated pulse profile in a statistical manner \citep{2014MNRAS.443.1463S}. For a simple case 
in which we assume the pulse profile is Gaussian shape and the phase jitter follows a
Gaussian distribution, the RMS error caused by jitter noise can be written 
as \citep{2010arXiv1010.3785C} 
\begin{align}\label{eq:jitter1}
\sigma_{\text{J}} = 0.28\,\mu s \left(\frac{W_i}{1\,\text{ms}}\right)\left(\frac{N}{10^6}\right)^{-1/2}
\left(\frac{f_{\text{J}}}{1/3}\right)\left(\frac{1+m_{\text{I}}^2}{2}\right)^{1/2} \,.
\end{align}
Here, $m_{\text{I}}$ is the amplitude modulation index defined as the ratio of 
the standard deviation of the amplitude to the mean at different pulse phases and 
$m_{\text{I}}\approx 1$. $f_{\text{J}}$ is the dimensionless jitter parameter 
defined as the ratio of the standard deviation of the phase of single pulse to 
the intrinsic pulse width $W_i$ of the template and $f_{\text{J}}\approx 1/3$. 
The RMS of total error $\sigma_{\text{t}}$ for estimated TOA is the quadratic 
summation of radiometer noise and jitter noise, i.e. $\sigma_{\text{t}}^2=\sigma_{\text{SNR}}^2+\sigma_{\text{J}}^2$.

By equating  Eq.~\ref{eq:raderror1} with Eq.~\ref{eq:jitter1}, we find that 
jitter noise will become more important than radiometer noise when $\text{SNR}_1$ 
exceeds only a few tenths. This sensitivity is accessible by the future radio telescopes such as 
FAST \cite{2014arXiv1407.0435H} and SKA \cite{2009A&A...493.1161S} that have 
larger collecting areas and lower system temperatures. In this 
scenario, the noise will not be reduced by increasing the observation bandwidth. 
Increasing the observation time will become an inevitable choice. As a result, PTAs will 
need to request more observation time of radio telescopes.

\subsection{Timing noise}

Timing noise, also known as spin noise, appears as the structures with 
temporal correlation in 
timing residuals that depart greatly from the measurement error alone. 
It may be caused by the irregularity of rotational spin rate which could root from 
the changes in internal structure and/or magnetosphere of neutron star. 
It has been found in a number of canonical pulsars and a few MSPs 
(e.g., B1937+21, B1821-24) \cite{2010MNRAS.402.1027H}. The RMS error of 
timing noise can be characterized by a scaling law \cite{2010ApJ...725.1607S}: 
\begin{align}\label{eq:TN1}
\sigma_{\text{TN}} = C\nu^{\alpha}|\dot{\nu}|^{\beta}T^{\gamma} \,, 
\end{align}
where $C$, $\alpha$, $\beta$ and $\gamma$ are the fitting parameters determined from 
the whole populations of pulsars with measurable timing noise or upper limits, 
$\nu$ and $\dot{\nu}$ is the pulsar spin frequency and frequency derivative, 
$T$ is the total span of observations. The best-fit values and $\pm 2\sigma$ confidence 
limits calculated from canonical pulsar and MSP population are $\ln C=1.6\pm 0.4$, $\alpha=-1.4\pm 0.1$, 
$\beta=1.1\pm 0.1$, and $\gamma=2.0\pm 0.2$ \cite{2010ApJ...725.1607S}. This suggests that MSPs 
with high spin frequency and low spin frequency derivative would have a lower level 
of timing noise. Eq.~\ref{eq:TN1} can serve as a criterion to select excellent 
timers that potentially have less timing noise from newly discovered MSPs for the PTAs. Despite the 
timing noise is not yet detected in most MSPs, it can be a latent phenomenon that may emerge 
from future data with longer observation span and higher timing precision. 

Timing noise is the red noise intrinsic to pulsar. For $\gamma \approx 2$, it 
has a power spectral index between $-4$ and $-6$  based on the random walk model 
of pulsar rotation. As a comparison, the characteristic strain $h_c$ of the 
stochastic GW background scales as $h_c = A (f/\text{yr}^{-1})^{\alpha}$, 
where $\alpha = -2/3$ for a background generated by the incoherent superposition 
of supermassive black hole binaries. The RMS residuals induced by this background 
grows as $T^{5/3}$ associated with a spectral index of $2\alpha-3=-13/3$ \cite{2006ApJ...653.1571J}. 
One of the consequences of the red noise is that the time of GW detection 
predicted by white noise dominated model \citep{2005ApJ...625L.123J} may 
be delayed up to 10 years depending on the level of red noise 
\citep{2013CQGra..30v4015S}. Besides, more MSPs will be required than 
previously expected in order to distinguish between the red noise from 
the GW stochastic background and the other red noise sources by cross-correlating 
the data from different pulsars, and the number of pulsars becomes a more important factor 
than the observation cadence and timing quality of individual pulsar when the PTA enters 
a regime where the lowest frequencies of the timing residuals are dominated by 
GWs, a likely case for the current PTAs \citep{2013CQGra..30v4015S}. 
Currently, the NANOGrav are adding 3-4 new MSPs discovered from the ongoing major pulsar surveys at 
Arecibo Observatory and Green Bank Telescope (e.g., PALFA and GBNCC) 
in the observation campaign each year.

\section{Noise assessment}

As discussed above, the characterization of noise in PTA plays a central 
role in hunting for GWs. Due to its importance, the NANOGrav 
has formed a noise budget working group to use complementary 
methods to assess the constituents of timing residuals and their 
physical causes. The white noise and Gaussian statistics are two 
essential aspects of the assessments, especially the latter is 
a common assumption in forming the GW detection strategies 
\cite{2012PhRvD..85d4034B, 2012ApJ...756..175E, 2013CQGra..30v4004E, 2014ApJ...795...96W}. 
Blindly applying these strategies without checking the presumption 
may lead to unreliable results. 

Methods used include autocorrelation analyses, Bayesian inference, 
zero-crossing tests, Gaussianity tests, etc. A memorandum to consolidate 
the results from the overall assessment of the NANOGrav 5-yrs ASP/GASP data 
set for 17 MSPs \cite{2013ApJ...762...94D} is in preparation. Further 
studies extended to the NANOGrav 9-yrs (including 4-yrs PUPPI/GUPPI) data 
set for more than 30 MSPs will be also carried out once the data are available. 

Initial results show that for the 5-yrs data most of the pulsars are consistent with 
the white noise assumption \cite{2013arXiv1311.3693P,2014inPrepYWANG}, 
although it is possible that the red timing noise can appear 
in the 9-yrs observations which have longer span and higher precision. Different levels 
of departure from Gaussian statistics are shown in most of the pulsars 
\cite{2014inPrepYWANG}, it is suspected that 
the diffractive interstellar scintillation is the root cause. This 
suggests that the robust signal detection and characterization methods that are 
not sensitive to the non-Gaussianity should be implemented in GW data analysis.

\section{Conclusions}

Noise characterization and mitigation are the central issues in detecting GWs by 
pulsar timing arrays. In this article, we provide an overview of the features of 
radiometer noise,  phase jitter noise and timing noise as well as the efforts to reduce them. 

The radiometer noise is dominant in the current timing precision, 
it will be continuously mitigated with the developments of instrument 
and algorithm to a level smaller than the jitter noise. 
The jitter noise is intrinsic to a pulsar, it can only be mitigated 
by extending integration time, therefore affects the strategies on 
telescope time application and allocation. 

Timing noise is latent for most MSPs, it can potentially postpone the detection of GWs by PTAs. 
Finding more MSPs with excellent timing performance in the 
ongoing and future surveys is imperative in the competition between the red noise 
from GWs and pulsars themselves.

\ack

We wish to acknowledge the invitation from Prof.~Andrea Lommen for 
the 10th LISA symposium, and the assistance from the local organizer 
at University of Florida, especially Prof.~Guido Mueller for his help in 
conference registration. We would like to thank the NANOGrav members for 
helpful discussion and the Center for Gravitational Wave Astronomy 
at UTB for partial support under NASA grant NNX09AV06A. The NANOGrav project 
is supported by the National Science Foundation under PIRE 
award number 0968296. 



\providecommand{\newblock}{}

\end{document}